\documentclass[pra,twocolumn,superscriptaddress,floatfix,amsmath,nofootinbib,amssymb]{revtex4-1}

\usepackage{graphicx}	
\usepackage{dcolumn}	
\usepackage{bm}		
\usepackage{color}
\usepackage{float}
\usepackage{mathtools}
\usepackage[]{caption}
\usepackage{subcaption}
\usepackage[pdf]{pstricks}
\usepackage{epstopdf}

\newcommand{\abs}[1]{\left| #1 \right|} 					
\newcommand{\ket}[1]{\left| #1 \right>} 					
\newcommand{\bra}[1]{\left< #1 \right|} 					
\newcommand{\diff}[1]{\ensuremath{\operatorname{d}\!{#1}\ }}	
\newcommand{\di}[1]{\ensuremath{\operatorname{d}\!{#1} }}	
\newcommand{\bb}{\mathbf} 							
\newcommand{\mc}{\mathcal} 							
\renewcommand{\Re}{\mathop{\rm Re}}					
\renewcommand{\Im}{\mathop{\rm Im}}					

\newcommand{\beq}{\begin{equation}}
\newcommand{\eeq}{\end{equation}}
\newcommand{\nn}{\nonumber}

\begin{document}

\title{Looking Inside a Black Hole\\}

\author{Alexander R. H. Smith}
\email{a14smith@uwaterloo.ca}
\affiliation{Department of Physics \& Astronomy, University of Waterloo, Waterloo, Ontario Canada N2L 3G1}

\author{Robert B. Mann}
\email{rbmann@uwaterloo.ca}
\affiliation{Department of Physics \& Astronomy, University of Waterloo, Waterloo, Ontario Canada N2L 3G1}

\date{\today}	

\begin{abstract}
The cosmic censorship conjecture posits that singularities forming to the future of a regular  Cauchy  surface are hidden by an event horizon. Consequently   any topological structures will ultimately collapse within the horizon of a set of black holes and so no observer can actively probe them classically.
We consider here a quantum analog of this problem, in which we compare the transition rates of   an Unruh-DeWitt detector  placed outside the horizon of an eternal BTZ black hole and its associated geon counterpart.  We find the transition rates differ, with the latter being time-dependent,
 implying that we are indeed able to probe the structure of the singularity from outside the Killing horizon.
\end{abstract}

\maketitle


The observed triviality of the spatial topology of our universe (namely its continuous deformability to $\bb R^3$) stands in stark contrast to the properties of general relativity, which admits all possible topologies.  The  topological censorship theorem~\cite{Friedman:1993ty,Galloway:1999bp} resolves this conundrum insofar as it relegates all isolated topological structures (such as wormholes) inaccessible to observers by any classical experimental means insofar  as    causal curves beginning near past null infinity passing through the interior 
of an asymptotically AdS (or flat) spacetime (obeying physically reasonable conditions) detect no topological structure not also present in the boundary-at-infinity \cite{Galloway:1999bp,Galloway:1999br}. 
 
Here we  show that observers can probe the global topology of a spacetime via quantum mechanical means.  Specifically, we examine two different situations in (2+1) dimensions: an Unruh-DeWitt detector sitting outside a non-rotating BTZ black hole and the same detector sitting outside the associated $\mathbb{R}\bb{P}^2$ geon. In both cases the detector travels along the same trajectory: not rotating and remaining at a fixed distance from the horizon. The difference between the two spacetimes arises in the topological structure of the singularity since the metrics for each  are locally identical.
The inability to actively probe such topology classically \cite{Friedman:1993ty,Galloway:1999bp}  naively suggests  identical transition rates, and an investigation of the $\mathbb{R}\bb{P}^3$ geon relative to its Schwarzschild counterpart  
led to a conjecture that the transition rates would be the same for  detector that only operates in the asymptotic past/future, as well as for a detector far from the horizon for finite times \cite{Louko:1998dj}. However passive observation of topological structures (e.g. via a white hole) is not forbidden, and since
quantum-mechanically the detector response is determined by the state we might expect its response rate to be 
 affected  by the global properties of the spacetime.  Topological identifications in spacetimes with constant curvature have been shown to yield different detector response rates~\cite{Louko:1998qf,Langlois:2005if,Langlois:2005nf} and it is therefore reasonable to expect differing  responses  even though the spacetime geometries near the trajectories in question happen to be identical. We find
 the detector does  indeed measure a difference in the transition rate, yielding a probe of the topological structure of the singularity.


We consider an Unruh-Dewitt detector coupled to a scalar field $\phi(x)$, treated to first order in perturbation theory \cite{Birrell:1982, Hodfkinson:2012, Hodfkinson:2012a}. The detector is a simplified model of a real particle detector that is linearly coupled to a scalar field. It can be thought of as an idealized atom with two energy levels, denoted by $\ket{0}_{d}$ and $\ket{E}_d$, with respective energy eigenvalues $0$ and $E$. 
The detector's interaction with the  field is described by the Hamiltonian
\beq
H_{\rm int} = c \chi(\tau) \mu(\tau) \phi(x(\tau)),
\eeq
where the point-like detector is moving along a timelike trajectory described by $x^{\mu}(\tau)$, 
$c$ is a small coupling constant,  $\mu(\tau)$ is the monopole moment of the detector, and $\chi(\tau)$ is a switching function that is positive during the interaction of the detector with the field and vanishing elsewhere. 

In general, as the detector moves along the trajectory $x^{\mu}(\tau)$, it will transition to $\ket{E}_d$ and the field will transition to some state $\ket{\psi}$. For $c<<1$, the probability of this transition to first order in perturbation theory is \cite{Hodfkinson:2012}
\beq
P(E) = c^2 \abs{ \bra{0_d} \mu(0) \ket{E_d} }^2 \mc{F}(E), \label{prob}
\eeq
where $\mc{F}(E)$ is the response function that encodes the information about the detector's trajectory, the switching function, and the initial state of the field. Explicitly  
\beq
\mc{F}(E) = \Re \int_{-\infty}^{\infty} \diff{u} \chi(u) \int_{0}^{\infty} \diff{s} \chi(u-s)e^{-iEs} G^+(u,u-s),
\eeq
where $G^+(x, x')$ is the Wightman Green function associated with the scalar field $\phi$.

The transition rate in the sharp switching limit is 
\beq
\dot{\mc{F}}(E) = \frac{1}{4} +2 \Im \int_{0}^{\Delta \tau} \diff{s}  e^{-iEs} G^+(\tau,\tau-s), \label{transitionrate}
\eeq
where $\Delta \tau =  \tau-\tau_0$ with $\tau_0$ being the proper time at which the detector was switched on, $\tau$ being the read-off time (the proper time at which the instantaneous transition rate is read off).


We wish  to compute the transition rate for a detector in the background of the BTZ black hole \cite{Banados:1992} and compare this to the associated geon  \cite{Louko:1999xb,Louko:2005}.  These solutions
can   be constructed by making appropriate  identifications of 3 dimensional anti-de Sitter (AdS$_3$) spacetime.  Specifically, AdS$_3$ can be obtained from the flat space $\mathbb{R}^{2,2}$ with coordinates $(X_1, X_2, T_1, T_2)$ and metric 
\beq
\di s^2 = - \di T_1^2 - \di T_2^2 +  \di X_1^2 +  \di X_2^2,
\eeq
by restriction  to the submanifold
\beq
X_1^2 - T_1^2 + X_2^2 - T_2^2 =  -\ell^2.
\eeq
with cosmological constant $\Lambda = -\ell^{-2}$.
The metric is
\begin{eqnarray}
\di s &=& - f(r) \di t^2 +\di r^2/f(r) + r^2 \di \phi^2  \label{SchwarzschildBTZ}\\
&=&- \frac{\ell^2}{(1+UV)^2} \left[ -4  \di U \di V + M \left(1-UV\right)^2 \di \phi^2 \right] \nonumber
\end{eqnarray}
where $f(r) = -M +\frac{r^2}{\ell^2}$; the  horizon is at $r_h = \ell \sqrt{M}$.  The relationship between each of the coordinates is (in the region where $T_1> \abs{X_1}$):
\begin{align}
T_1 &= \ell \left(\frac{1-UV}{1+UV}\right) \cosh \sqrt{M} \phi =\ell \sqrt{\alpha(r)} \cosh \left( \frac{r_h}{\ell} \phi \right),  \nn \\
X_1 &= \ell \left(\frac{1-UV}{1+UV}\right) \sinh \sqrt{M} \phi=  \ell \sqrt{\alpha(r)} \sinh \left( \frac{r_h}{\ell} \phi \right), \nn \\
T_2 &= \ell \left(\frac{V+U}{1+UV}\right)=\ell \sqrt{\alpha(r)-1} \sinh \left( \frac{r_h}{\ell^2} t \right), \nn \\
X_2 &= \ell \left(\frac{V-U}{1+UV}\right) = \ell \sqrt{\alpha(r)-1} \cosh \left( \frac{r_h}{\ell^2} t \right), \label{X2}
\end{align}
where $\alpha(r) = r^2/r_h^2$, $-1<UV<1$ and $M>0$. The coordinates $U$ and $V$ are null coordinates 
and the BTZ spacetime is  obtained by making a periodic identification in the coordinate $\phi$: $(U, V, \phi) \sim (U, V, \phi + 2 \pi)$.

\begin{figure}[H]
\centering
\scalebox{0.95}{
\begin{subfigure}[b]{0.3 \textwidth}
\centering
\begin{pspicture}(5,5)
	\psline{-}(0,0)(0,5) 			
	\psline{-}(5,0)(5,5)			
	\psline{-}(0,5)(5,0)			
	\psline{-}(0,0)(5,5)			
	\psline[linestyle=dashed](2.5,0.08)(2.5,5) 
	\pscurve[showpoints=false](0.,0.)(0.1,0.08)(0.2,0.)(0.3,-0.08)(0.4,0.) 	
	(0.5,0.08)(0.6,0.)(0.7,-0.08)(0.8,0.)(0.9,0.08)(1.,0.)(1.1,-0.08)(1.2,0.)
	(1.3,0.08)(1.4,0.)(1.5,-0.08)(1.6,0.)(1.7,0.08)(1.8,0.)(1.9,-0.08)(2.,0.)
	(2.1,0.08)(2.2,0.)(2.3,-0.08)(2.4,0.)(2.5,0.08)(2.6,0.)(2.7,-0.08)(2.8,0.)
	(2.9,0.08)(3.,0.)(3.1,-0.08)(3.2,0.)(3.3,0.08)(3.4,0.)(3.5,-0.08)(3.6,0.)
	(3.7,0.08)(3.8,0.)(3.9,-0.08)(4.,0.)(4.1,0.08)(4.2,0.)(4.3,-0.08)(4.4,0.)
	(4.5,0.08)(4.6,0.)(4.7,-0.08)(4.8,0.)(4.9,0.08)(5.,0.)
	\pscurve[showpoints=false](0., 5.)(0.1, 5.08)(0.2, 5.)(0.3, 4.92)(0.4, 5.)	
	(0.5, 5.08)(0.6, 5.)(0.7, 4.92)(0.8, 5.)(0.9, 5.08)(1., 5.)(1.1, 4.92)(1.2, 5.)
	(1.3, 5.08)(1.4, 5.)(1.5, 4.92)(1.6, 5.)(1.7, 5.08)(1.8, 5.)(1.9, 4.92)(2., 5.)
	(2.1, 5.08)(2.2, 5.)(2.3, 4.92)(2.4, 5.)(2.5, 5.08)(2.6, 5.)(2.7, 4.92)(2.8, 5.)
	(2.9, 5.08)(3., 5.)(3.1, 4.92)(3.2,5.)(3.3, 5.08)(3.4, 5.)(3.5, 4.92)(3.6, 5.)
	(3.7, 5.08)(3.8, 5.)(3.9, 4.92)(4., 5.)(4.1, 5.08)(4.2, 5.)(4.3, 4.92)(4.4, 5.)
	(4.5, 5.08)(4.6, 5.)(4.7, 4.92)(4.8, 5.)(4.9, 5.08)(5., 5.)
\end{pspicture}
\caption{}
\label{fig:BTZconformal}
\end{subfigure}}%
          \quad
\scalebox{0.95}{
\begin{subfigure}[b]{0.15 \textwidth}
\centering
\begin{pspicture}(5,2.5)
	\psline[linestyle=dashed](0,0)(0,5) 	
	\psline(2.5,0.08)(2.5,5.08) 		
	\psline{-}(0,2.5)(2.5,5)			
	\psline{-}(2.5,0)(0,2.5)			
	\pscurve[showpoints=false](0.,0.)(0.1,0.08)(0.2,0.)(0.3,-0.08)(0.4,0.) 	
	(0.5,0.08)(0.6,0.)(0.7,-0.08)(0.8,0.)(0.9,0.08)(1.,0.)(1.1,-0.08)(1.2,0.)
	(1.3,0.08)(1.4,0.)(1.5,-0.08)(1.6,0.)(1.7,0.08)(1.8,0.)(1.9,-0.08)(2.,0.)
	(2.1,0.08)(2.2,0.)(2.3,-0.08)(2.4,0.)(2.5,0.08)
	\pscurve[showpoints=false](0., 5.)(0.1, 5.08)(0.2, 5.)(0.3, 4.92)(0.4, 5.)	
	(0.5, 5.08)(0.6, 5.)(0.7, 4.92)(0.8, 5.)(0.9, 5.08)(1., 5.)(1.1, 4.92)(1.2, 5.)
	(1.3, 5.08)(1.4, 5.)(1.5, 4.92)(1.6, 5.)(1.7, 5.08)(1.8, 5.)(1.9, 4.92)(2., 5.)
	(2.1, 5.08)(2.2, 5.)(2.3, 4.92)(2.4, 5.)(2.5, 5.08)
\end{pspicture}
\caption{}
\label{fig:geonConformal}
\end{subfigure}}
     \caption{ (a) A conformal diagram of the BTZ hole. Each point in the diagram represents a suppressed $S^1$. (b) A conformal diagram of the geon associated with the BTZ spacetime. The region not on the dashed line is identical to the diagram in (a); however on the dashed line each point in the diagram again represents a suppressed $S^1$ but with half the circumference of the suppressed $S^1$ in diagram (a).}
    \label{fig:conformal}
\end{figure}
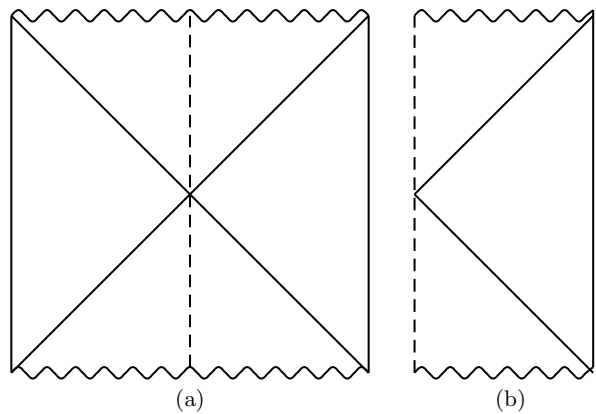

To obtain the geon quotient  of the BTZ spacetime  \cite{Louko:2005} we  note  from Eq. \eqref{SchwarzschildBTZ} that the metric admits the Killing vector $\xi = \partial_{\phi}$, which generates the freely-acting involutive isometry $P = \exp{ \left(\pi r_h \xi\right)}$. The map
\beq
J: \left(U, V, \phi \right) \rightarrow \left(V, U, P(\phi)\right), \label{Jisometry}
\eeq
is then a freely-acting involutive isometry acting on the entire BTZ spacetime $\mathcal{M}_{\rm BTZ}$. Together with the identity map it generates the isometry group $\Gamma := \left\{ {\rm Id}_{\rm BTZ}, P \right\} \simeq \mathbb{Z}_2$.
The geon spacetime $\mathcal{M}_{\rm geon}$ is then the quotient spacetime of the BTZ hole with the group $\mathbb{Z}_2$:
\beq
\mathcal{M}_{\rm geon} = \mathcal{M}_{\rm BTZ} / \mathbb{Z}_2.
\eeq
 It is identical to the BTZ spacetime outside the horizon, but (see Fig. \ref{fig:conformal}) inside  the topology of the singularity changes from that of a point to the real projective space $\mathbb{R}\mathbf{P}^2$   \cite{Louko:2005}.  Likewise, the
Hartle-Hawking vacuum $\ket{0_K}$, which is the vacuum of mode functions of positive frequency with respect to the affine parameters of the horizon-generating null geodesics, is invariant under the involution $J$. Hence it induces a unique vacuum on the  geon \cite{Louko:1998dj}, which we denote by $\ket{0_G}$.

To obtain the transition rate of a detector in the BTZ spacetime we require the appropriate Wightman function as demanded by Eq. \eqref{transitionrate}. The Wightman functions of the BTZ spacetime are well-known and obtained via the method of images \cite{Carlip:1995, Hodfkinson:2012, Hodfkinson:2012a}:  
\beq
G_{\rm BTZ }(x,x') = \sum_n G^{(\zeta)}_A\left(x, \Lambda^nx'\right), \label{BTZWightmann}
\eeq
where $\Lambda x'$ denotes the periodic identification $(t,r,\phi) \sim (t,r,\phi+2\pi)$ on the point $x'$ and the function $G_A^{(\zeta)}$ denotes the AdS$_3$ Wightman function
\begin{align}
G_A^{(\zeta)}(x,x') =& \frac{1}{4\pi}\left(\frac{1}{\sqrt{\Delta X^2(x,x')}} \right. \nn \\
&\qquad \qquad \left. - \frac{\zeta}{\sqrt{\Delta X^2(x,x')+4\ell^2}}\right), \label{BTZWightmann2}
\end{align}
where the parameter $\zeta$ specifies the boundary condition at infinity --- either transparent, Dirichlet, or Neumann corresponding to $\zeta$ taking the values $\left\{0,1,-1\right\}$ respectively --- and the function $\Delta X^2(x,x')$ is the squared geodesic distance between $x$ and $x'$ in the flat $\mathbb{R}^{2,2}$ space given by
\begin{align}
\Delta X^2(x,x') =& -\left(T_1 - T_1'\right)^2-\left(T_2 -T_2'\right)^2 \nn \\
&+\left(X_1 -X_1'\right)^2 +\left(X_2 -X_2'\right)^2.
\end{align}

The transition rate of the detector is obtained by substituting Eq. \eqref{BTZWightmann} into Eq. \eqref{transitionrate}, which yields
\begin{align}
\dot{\mathcal{F}}_{\tau}^{\rm BTZ} (E) =& \frac{1}{4} + \frac{1}{2 \pi \sqrt{2}} \sum_{n=-\infty}^{\infty} \int_0^{\Delta \tau/ \ell} \diff{ \tilde{s}} \nn \\
&\Re \left[e^{-iE\ell \tilde{s}} \left( \frac{1}{\sqrt{\Delta \tilde{X}^2_n}} -  \frac{\zeta}{\sqrt{\Delta \tilde{X}^2_n+2}}\right)\right],
\end{align}
where we have introduced the dimensionless integration variable $\tilde{s} := s/ \ell$ and written
\begin{align}
\Delta \tilde{X}_n^2 	:=& \Delta X^2 \left(x(\tau), \Lambda^n x(\tau- \ell \tilde{s})\right)  \\
				=&-1 + \sqrt{\alpha(r) \alpha(r')}\cosh \left[(r_h/\ell) (\phi-\phi'-2\pi n )\right]  \nn \\
				&- \sqrt{(\alpha(r)-1)( \alpha(r')-1)}\cosh \left[(r_h/\ell)(t-t')\right] \nn
\end{align}
where the unprimed coordinates are evaluated at $x(\tau)$ and the primed coordinates are evaluated at $x(\tau - \ell \tilde{s})$.


Let us suppose the trajectory of our detector is such that it is sitting at a constant distance from the centre of the black hole so that the trajectory is given by
\beq
r = {\rm constant}, \ \ \ \ t = \frac{ \ell }{\sqrt{r^2 - r_h^2}} \  \tau, \ \ \ \phi = 0, \label{trajectory}
\eeq
where $r$ specifies the radial location and $\tau$ is the proper time of the detector. $\Delta \tilde{X}_n^2$ becomes
\beq
\Delta \tilde{X}_n^2 	= \ 2 \left(\alpha -1 \right) \left[  K_n- \sinh^2(\Xi \tilde{s}) \right],
\eeq
where we have suppressed the dependence of $\alpha$ on $r$ and we have introduced $K_n := (1-\alpha^{-1})^{-1} \sinh^2(n\pi r_+ /\ell)$ and $\Xi := (2\sqrt{\alpha-1})^{-1}$. The transition rate is now 
\begin{align}
\dot{\mathcal{F}}^{\rm BTZ} (E)  = \frac{1}{4} + \frac{1}{4\pi \sqrt{\alpha-1}} \sum_{n=-\infty}^{\infty} \int_0^{\infty} \diff{ \tilde{s}} \Re \left[ e^{-iE\ell \tilde{s}}\right. \nn \\
\left.\left( \frac{1}{\sqrt{K_n - \sinh^2\left(\Xi \tilde{s} \right)}} -\frac{\zeta}{\sqrt{Q_n - \sinh^2\left(\Xi \tilde{s} \right)}}\right)\right], \label{BTZTransition1}
\end{align}
where $Q_n :=  K_n+(\alpha -1)^{-1}$  and we have taken the switch on time of the detector to be in the asymptotic past so that $\Delta \tau/\ell$ goes to infinity. 

This expression can be further simplified to   \cite{Hodfkinson:2012}
\begin{align}
\dot{\mathcal{F}}^{\rm BTZ} (E)  = \frac{e^{-\beta E \ell /2}}{2\pi} \sum_{n=-\infty}^{\infty} \int_0^{\infty} \di y \cos\left(y \beta E \ell / \pi\right) \nn \\
 \left[  \frac{1}{\sqrt{K_n + \cosh^2 y }} -\frac{\zeta}{\sqrt{Q_n + \cosh^2y}}\right], \label{BTZTransition2}
\end{align}
where $\beta = \pi \Xi^{-1} = 2 \pi \sqrt{\alpha-1}$ and   $y=\Xi \tilde{s}=\frac{\pi}{\beta} \tilde{s}$.
We have  dropped the subscript $\tau$ since the transition rate does not depend on  the  read-off time of the detector. This will not be true in the case of the geon.
  
Since the geon spacetime is obtained by acting the isometry $J$, Eq. \eqref{Jisometry}, on the BTZ spacetime, we can again make use of the method of images to obtain the geon Wightman functions
\begin{align}
G_{\rm geon }(x,x') &= \sum_{m\in\{0,1\}} G^{(\zeta)}_{\rm BTZ} \left(x, J^mx'\right) \nn \\
				&= \sum_{m\in\{0,1\}}\sum_n G^{(\zeta)}_A \left(x, J^m \Lambda^n x'\right),
\end{align}
where we only sum over $m\in\{0,1\}$ since $J$ is an involutive isometry. The $m=0$ term is identical to Eq. \eqref{BTZWightmann2}. The $m=1$ term is obtained by applying $J$ once to the primed coordinates. Swapping $U$ and $V$ results in $t \rightarrow -t$ and  $X_2 \rightarrow -X_2$ as can be seen from Eqs.  \eqref{X2}.  Applying $P(\phi)$ we have $\phi' \rightarrow \phi' + \pi$; 
substituting this into Eq. \eqref{transitionrate} yields 
\beq
\dot{\mathcal{F}}^{\rm geon}_{\tau}(E) = \dot{\mathcal{F}}^{\rm BTZ} (E) + \Delta\dot{\mathcal{F}}^{\rm geon}(E, \tau) , \label{GeonTransition}
\eeq
for the transition rate of a detector in the geon space-time traversing  the same trajectory described in Eq. \eqref{trajectory}.

\begin{figure}[H]
        \centering
        \scalebox{0.9}{
        \begin{subfigure}[b]{0.42\textwidth}
                \centering
                \includegraphics[width=\textwidth]{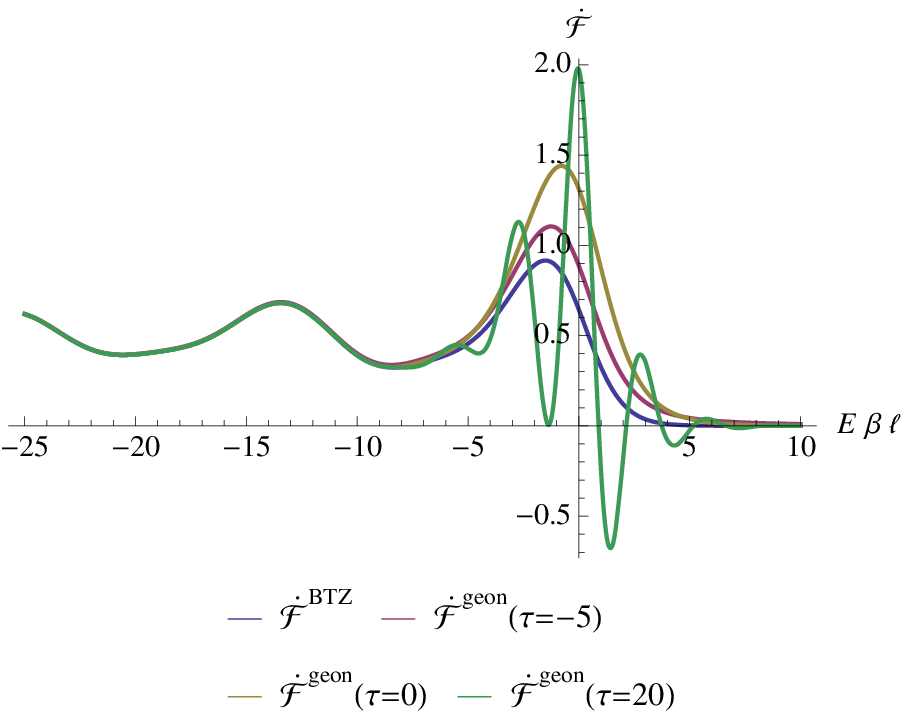}
                \caption{}
                \label{fig:TransitionRateE2}
        \end{subfigure}}
        \scalebox{0.9}{
        \begin{subfigure}[b]{0.42\textwidth}
                \centering
                \includegraphics[width=\textwidth]{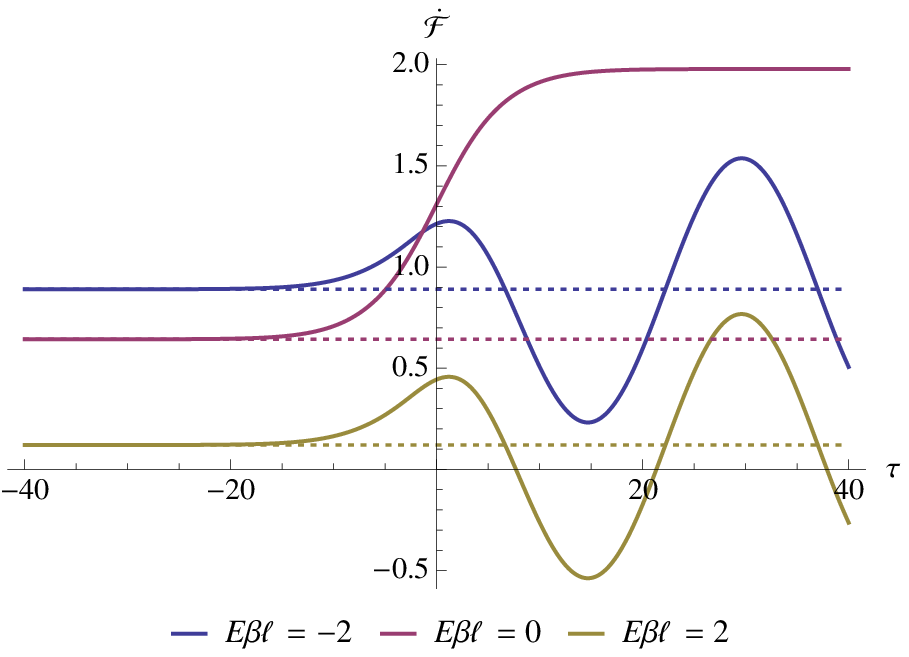}
                \caption{}
                \label{fig:TransitionRateTau2}
        \end{subfigure}
        }
        \caption{The transition rate of a stationary detector in both the BTZ and geon spacetimes as a function of (a) $E \beta \ell$ and (b) the proper time of the detector $\tau$ at which the detector is read, with $r_h/ \ell = 0.5$, $\alpha = 10$, and $\zeta=0$; the dotted line is the ($\tau$-independent) BTZ transition rate.}
        \label{fig:Numerics}
\end{figure}

The additional contribution due to the geon is 
\begin{align}
&\Delta\dot{\mathcal{F}}^{\rm geon}(E, \tau)
  =   \frac{1}{2\pi }  \sum_{n=-\infty}^{\infty} \int_0^{\infty} \diff{ y}   \cos (E\ell \beta y / \pi)  \label{geonAddition} \\ 
&\left( \frac{1}{\sqrt{\bar{K}_{n} + \cosh^2\left(y - 2 \tau \Xi \right)}} -\frac{\zeta}{\sqrt{\bar{Q}_{n} + \cosh^2\left( y - 2 \tau \Xi  \right)}} \right) \nn
\end{align}
with
\begin{align}
\bar{K}_{n} &:= (1-\alpha^{-1})^{-1} \sinh^2\left(\left(n+\tfrac{1}{2}\right)\pi r_+ /\ell\right),  \nn\\
\bar{Q}_{n} &:=  \bar{K}_{n}+(\alpha -1)^{-1}. \nn
\end{align}
 We see that the transition rate of the detector in the geon spacetime now depends on the transition rate read-out time $\tau$. 


Having now derived the transition rate of a detector in the BTZ spacetime, Eq. \eqref{BTZTransition2}, and in the associated geon spacetime, Eq. \eqref{GeonTransition}, we are able to evaluate the rates numerically, as shown in Fig. \ref{fig:Numerics}. The BTZ transition rate depends on 4 parameters: the energy gap of the detector $E$, which appears only in the dimensionless ratio $E \beta \ell$; the horizon radius $r_h =  \ell \sqrt{M}$, which characterizes the mass of the hole; $\alpha(r)$ which encodes the trajectory of the detector; and $\zeta$ which specifies the boundary conditions the AdS$_3$ Wightman function satisfies.  The geon transition rate depends on the same 4 parameters and the additional parameter $\tau$, the time at which the transition rate is read off of the detector.

The sums appearing in Eqs. \eqref{BTZTransition2} and \eqref{GeonTransition} were evaluated for $-3\leq n \leq 3$ as larger terms do not contribute significantly to the transition rate. This can be seen by noting that for the $n$th term, the denominator appearing in the integrand goes as $ e^n$. The integrals appearing in the transition rates were evaluated using Gauss-Kronrod quadrature \cite{Davis:1984} to a precision of $\pm 0.001$. 

\begin{figure}[H]
    \centering
    \includegraphics[width=0.4\textwidth]{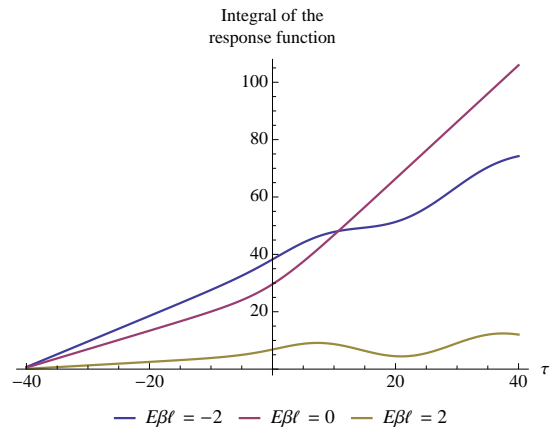}
    \caption[justification=justified]{ A plot of the integral of the derivative of the response function (the integral of the curves given in Fig. \ref{fig:Numerics} (b)) as a function of $\tau$. From Eq. \eqref{prob} this is proportional to the probability, which must remain positive. A hundred points were uniformly sampled between $-40$ and $40$ and a simple Riemann sum integration method was used.   \label{fig:integral}}
\end{figure}

We  see  from Fig. \ref{fig:Numerics} that for increasingly negative $\tau$ the  BTZ and geon transition rates become virtually identical, whereas for increasingly positive $\tau$ they significantly differ, with the geon  transition rate becoming increasingly oscillatory.   We also see that for sufficiently large $E$ the transition rate is sometimes negative.  The interpretation of this requires some care \cite{Langlois:2005if,Langlois:2005nf}. While the transition rate is proportional to the derivative of the probability of the transition with respect to the switch off time, this is not the same thing as the response of a detector turned on at time $\tau$ and then turned off at time $\tau+\delta\tau$ in the limit $\delta\tau \to 0$.  Rather it corresponds to the comparison of detector responses in two different ensembles of detectors, one switched off at $\tau$ the other switched off at $\tau+\delta\tau$ as $\delta\tau \to 0$.  The overall transition probability  remains positive, as illustrated in Fig. \ref{fig:integral} for the geon.
 

In this paper we have shown that a detector placed outside a BTZ hole and the same detector placed outside the associated geon will have different transition rates, even though the local geometry is identical, yielding a probe of the the topology of a spacetime via local nonstationary measurements even if the topological structure is hidden behind a Killing horizon. 
It is reasonable to expect that similar phenomena hold in higher dimensions. Whether or not some generalized quantum version of the (active) topological censorship theorem holds remains an interesting subject for future study.
\\
\\
{\it Acknowledgements}  This work was supported in part by the Natural Sciences and Engineering Research Council of Canada and an Ontario Graduate Scholarship.   We are grateful to Jorma Louko for   helpful discussions, and to Don Marolf, Kristin Schleich, and Don Witt for helpful correspondence.
\\

\vspace{-1.0cm}
\bibliographystyle{unsrt}
\bibliography{TopologicalCensorship}

\end{document}